\documentclass[final,5p,times,twocolumn]{elsarticle}

\usepackage{amsmath}
\usepackage{amssymb}
\usepackage{graphicx}
\usepackage{bm}
\usepackage{braket}
\usepackage{enumitem}
\usepackage{tensor}
\usepackage{color}
\usepackage{combelow}
\usepackage{slashed}
\usepackage{epstopdf}

\def\omegat{\widetilde{\omega}}

\def\omegabar{\overline{\omega}}

\journal{Physics Letters B}

\begin{document}

\begin{frontmatter}

\title{Quantum non-equilibrium effects in rigidly-rotating thermal states}

\author[wut]{Victor E. Ambru\cb{s}}
\ead{victor.ambrus@e-uvt.ro}
\address[wut]{Department of Physics, West University of Timi\cb{s}oara,
Bd.~Vasile P\^arvan 4, Timi\cb{s}oara, 300223, Romania}

\date{\today}

\begin{abstract}
Based on known analytic results, the thermal expectation value of the 
stress-energy tensor (SET) operator for the massless Dirac field is analysed from a hydrodynamic 
perspective. Key to this analysis is the Landau decomposition of the SET, with the aid of 
which we find terms which are not present in the ideal SET predicted by kinetic theory.
Moreover, the quantum corrections become dominant in the vicinity of the speed of light 
surface (SOL). While rigidly-rotating thermal states cannot be constructed for the Klein-Gordon 
field, we perform a similar analysis at the level of quantum corrections previously reported 
in the literature and we show that the Landau frame is well-defined only when the system 
is enclosed inside a boundary located inside or on the SOL.
We discuss the relevance of these results for accretion disks around rapidly-rotating pulsars.\\
{\footnotesize 
\copyright 2017. This manuscript version is made available under the CC-BY-NC-ND 4.0 license 
http://creativecommons.org/licenses/by-nc-nd/4.0/.
}
\end{abstract}


\end{frontmatter}

\section{Introduction} \label{sec:intro}

In relativistic fluid dynamics, global thermal equilibrium can be attained 
if the product $\beta u^\mu$ between the inverse local temperature $\beta$ and 
the four-velocity $u^\mu$ of the flow satisfies the Killing equation 
\cite{degroot80,cercignani02,becattini12,becattini15epjc,becattini16appb}.
A special property of thermal equilibrium is that the stress-energy tensor (SET) 
$T^{\mu\nu}_{\rm eq} = (E + P) u^\mu u^\nu + P g^{\mu\nu}$ 
corresponds to that of an ideal fluid of energy density $E$ and pressure $P$
\cite{cercignani02,ambrus15wut,ambrus16kin,rezzolla13}.\footnote{We use Planck units 
with $c = \hbar = k_{B} = 1$, while the metric signature is $(-,+,+,+)$.} 
In this letter, we will show that a quantum 
field theory (QFT) computation of the SET for rigidly-rotating thermal states (RRTS)
contains non-ideal terms, as well as corrections to $E$ 
which become important near the speed of light surface (SOL). 
We discuss the relevance of these corrections in the context of an astrophysical application.

\section{Kinetic theory analysis} \label{sec:kin}

In space-times with axial symmetry, RRTS in thermal equilibrium 
can be described using the Killing vector corresponding to rotations about the 
$z$ axis, i.e., $\beta u = \beta_0(\partial_t + \Omega \partial_\varphi)$,
where $\Omega$ is the angular velocity of the rotating state \cite{ambrus16kin}.
On Minkowski space, the particle four-flow $N^\mu_{\rm eq}$ and stress-energy tensor 
$T^{\mu\nu}_{\rm eq}$ corresponding to RRTS are given by:
\begin{equation}\label{eq:macro_eq}
 N^\mu_{\rm eq} = n u^\mu, \qquad
 T^{\mu\nu}_{\rm eq} = (E + P) u^\mu u^\nu + P g^{\mu\nu},
\end{equation}
while $\beta$ and $u = u^\mu \partial_\mu$ are given by: 
\begin{equation}
 \beta = \gamma^{-1} \beta_0, \qquad 
 u = \gamma(\partial_t + \Omega \partial_\varphi),
 \label{eq:beta_u}
\end{equation}
where $\gamma$ is the Lorentz factor of a co-rotating observer at distance $\rho$ from the $z$ axis:
\begin{equation}
 \gamma = \left(1 - \rho^2 \Omega^2\right)^{-1/2}.
 \label{eq:gamma}
\end{equation}
The Killing vector $\beta u$ becomes null on the SOL, 
where $\rho\Omega \rightarrow 1$ and co-rotating observers travel at the speed of light.
From Eq.~\eqref{eq:beta_u}, it can be seen that the temperature $\beta^{-1}$ diverges 
as the SOL is approached. 
The energy density $E$ for massless particles obeying Fermi-Dirac (F-D) and 
Bose-Einstein (B-E) statistics is given by \cite{ambrus15wut}:
\begin{equation}
 E_{\rm F-D}= \frac{7\pi^2 \gamma^4}{60 \beta_0^4}, \qquad
 E_{\rm B-E} = \frac{\pi^2 \gamma^4}{30 \beta_0^4},\label{eq:E_kin}
\end{equation}
while $P = E/3$. Since $E$ and $P$ diverge as inverse powers of the distance to the SOL,
RRTS are well-defined only up to the SOL.
While such divergent states clearly cannot occur in nature, rigid rotation can be induced
in astrophysical systems, such as accretion disks around rapidly-rotating neutron stars or 
magnetars, where the intense magnetic field can lock charged particles into rigid 
rotation.\footnote{In such systems, various mechanisms prevent the violation of special 
relativity \cite{meier12}.} We investigate the role of quantum corrections in such systems 
in Sec.~\ref{sec:astro}.

\section{Stress-energy tensor decompositions} \label{sec:SET}

Before discussing the quantum analogue of Eqs.~\eqref{eq:E_kin}, the tools necessary to analyse 
the SET in out of equilibrium states must be introduced.
The main difficulty comes due to the equivalence between mass and energy in special relativity, 
which makes the distinction between the velocity $u^\mu$ and the heat flux $q^\mu$ ambiguous.
For a general (time-like) choice of $u^\mu$, $N^\mu$ can be decomposed as \cite{bouras10}:
\begin{equation}
 N^\mu = nu^\mu + V^\mu,
\end{equation}
where $n = -u_\mu N^\mu$ and the flow of particles in the local rest frame (LRF) $V^\mu$ is given by:
\begin{equation}
 V^\mu = \Delta^\mu{}_\nu N^\nu
\end{equation}
In the above, $\Delta^{\mu\nu} = u^\mu u^\nu + g^{\mu\nu}$
is the projector on the hypersurface orthogonal to $u^\mu$.
The decomposition of the SET reads:
\begin{equation}
 T^{\mu\nu} = E u^\mu u^\nu + (P + \omegabar) \Delta^{\mu\nu} + 
 W^\mu u^\nu + W^\nu u^\mu + \pi^{\mu\nu},\label{eq:SET}
\end{equation}
where the dynamic pressure $\omegabar$, 
flow of energy in the LRF $W^\mu$ and 
shear stress $\pi^{\mu\nu}$, together with $V^\mu$, represent non-equilibrium terms.
The quantities on the right hand side of Eq.~\eqref{eq:SET} 
can be obtained through:
\begin{gather}
 E = u^\mu u^\nu T_{\mu\nu}, \quad 
 P + \omegabar= \frac{1}{3} \Delta^{\mu\nu} T_{\mu\nu},\quad
 W^\mu = -\Delta^{\mu\nu} u^{\lambda} T_{\nu\lambda},\nonumber\\
 \pi^{\mu\nu} = \left(\Delta^{\mu\lambda} \Delta^{\nu\sigma} - 
 \frac{1}{3} \Delta^{\mu\nu} \Delta^{\lambda\sigma}\right) T_{\lambda\sigma},
 \label{eq:SET_dec}
\end{gather}
For a massless fluid, $\omegabar = 0$.
The heat flux $q^\mu$ is defined as \cite{bouras10}:
\begin{equation}
 q^\mu = W^\mu - \frac{E + P}{n} V^\mu.
\end{equation}

In the Eckart (particle) frame \cite{cercignani02,rezzolla13,eckart40}, 
$u^\mu_e$ is defined as the unit vector parallel to $N^\mu$. 
Observers in the LRF of the Eckart frame see a flow of energy ($W^\mu_e = q^\mu_e$) 
and no flow of particles ($V^\mu_e = 0$).
Since $N^\mu$ cannot be obtained using the QFT approach considered in this 
paper, the Eckart velocity $u^\mu_e$ 
cannot be defined. Hence, we will not consider the Eckart frame further
in this paper.

In the Landau (energy) frame \cite{cercignani02,rezzolla13,landau87},
$u^\mu \equiv u^\mu_L$ is defined as the eigenvector of $T^{\mu}{}_{\nu}$ 
corresponding to the (real, positive) Landau energy density $E_L$:
\begin{equation}
 T^\mu{}_\nu u_L^\nu = -E_L u_L^\mu,\label{eq:Landau_def}
\end{equation}
such that $W^\mu_L = 0$, which implies that there is no energy flux in the LRF.
Since $V^\mu_L = -\frac{n_L}{E_L + P_L} q^\mu_L$ is in general non-zero, 
an observer in the LRF of the Landau frame will detect a non-vanishing particle flux.

Finally, the $\beta$-frame (thermometer frame)
for the case of rigid rotation is defined with respect to \cite{becattini15epjc}:
\begin{equation}
 u_\beta = \gamma(\partial_t + \Omega \partial_\varphi).\label{eq:ubeta}
\end{equation}
A special property of the $\beta$-frame is that the LRF temperature is 
highest compared to the temperature measured with respect to any other frame \cite{becattini15epjc}.
In general, $V_\beta^\mu$ and $W_\beta^\mu$ do not vanish, such that 
the $\beta$-frame is a mixed particle-energy frame \cite{van14}.
Due to the simplicity of its construction, we will start the analysis of the 
quantum SET with respect to the $\beta$-frame.

\section{Klein-Gordon field} \label{sec:kg}

We now analyse the construction of RRTS from a QFT pers\-pective.
A first surprise comes from the analysis of the 
RRTS of the Klein-Gordon field: in the unbounded Minkowski space,
there exist modes which have a non-vanishing Minkowski energy $\omega$ (i.e., with respect to 
the static Hamiltonian $H_s = i\partial_t$), while their co-rotating energy $\omegat = \omega - \Omega m$,
measured with respect to the rotating Hamiltonian $H_r = i(\partial_t + \Omega \partial_\varphi)$,
vanishes. For such modes, the Bose-Einstein density of states factor $(e^{\beta\omegat} - 1)^{-1}$ 
diverges, yielding divergent thermal expectation values (t.e.v.s) 
at every point in the space-time \cite{ambrus14plb,duffy03}.
The kinetic theory result \eqref{eq:E_kin} 
is clearly unaffected by this vanishing co-rotating energy modes catastrophy.
Indeed, the problematic modes are no longer present in the QFT formulation if the system 
is enclosed within a boundary placed inside or on the SOL \cite{duffy03,nicolaevici01}. 
Furthermore, a recent perturbative QFT 
analysis allows the computation of quantum corrections to the kinetic theory SET \cite{becattini15}, 
which we will analyse in detail in what follows.
For completeness, we present an analysis of 
the connection between these perturbative results and the non-perturbative QFT approach 
in \ref{app:kg}.

Substituting the results in Table~III of Ref.~\cite{becattini15} into Eq.~(34) in Ref.~\cite{becattini15} 
yields the following $\beta$-frame \eqref{eq:beta_u} decomposition of the SET:
\begin{gather}
 E_{\beta} = \frac{\pi^2 \gamma^4}{30\beta_0^4} + 
 \frac{\Omega^2 \gamma^6}{36\beta_0^2}, \qquad
 W_{\beta} = \frac{\Omega^3 \gamma^7}{18\beta_0^2} 
 \left(\rho^2\Omega \partial_t + \partial_\varphi\right),\nonumber\\
 \pi^{\mu\nu}_{\beta} = \frac{\Omega^2 \gamma^6}{54\beta_0^2} 
 \begin{pmatrix}
  \gamma^2 - 1 & 0 & \Omega \gamma^2 & 0\\
  0 & 1 & 0 & 0\\
  \Omega \gamma^2 & 0 & \rho^{-2} \gamma^2 & 0\\
  0 & 0 & 0 & -2
 \end{pmatrix},
 \label{eq:KG_kin}
\end{gather}
where $\omega = \Omega \gamma^2 \partial_z$, $a = \rho\Omega^2 \gamma^2 \partial_\rho$ and 
$\gamma = \beta_0^2 \Omega^3 \gamma^3 (\rho^2\Omega \partial_t + \partial_\varphi)$ were 
used in Eq.~(34) of Ref.~\cite{becattini15}.
Compared to the kinetic theory result \eqref{eq:macro_eq}, the quantum SET contains 
non-vanishing contributions in the form of the non-ideal terms $W^\mu$ and $\pi^{\mu\nu}$.
Moreover, the second term in $E_{\beta}$ \eqref{eq:KG_kin}
represents a correction to $E_{\rm B-E}$ \eqref{eq:E_kin} which becomes dominant
in the vicinity of the SOL due to the $\gamma^6$ factor.

The construction of the Landau frame yields:
\begin{align}
 E_L =& \frac{E_{\beta}}{3} - \frac{1}{2}\bm{\hat{W}}_{\beta} \cdot \bm{\pi}_{\beta} \cdot \bm{\hat{W}}_{\beta} +
 \sqrt{\left(\frac{2E_{\beta}}{3} + \frac{1}{2}\bm{\hat{W}}_{\beta} \cdot \bm{\pi}_{\beta} \cdot \bm{\hat{W}}_{\beta}\right)^2 
 - W^2_{\beta}},\label{eq:kg_EL}\\
 u_L^\mu =& \sqrt{\frac{E_L + \frac{1}{3} E_{\beta} + \bm{\hat{W}}_{\beta} \cdot \bm{\pi}_{\beta} \cdot \bm{\hat{W}}_{\beta}}
 {2(E_L - \frac{1}{3}E_{\beta} + \frac{1}{2} \bm{\hat{W}}_{\beta} \cdot \bm{\pi}_{\beta} \cdot \bm{\hat{W}}_{\beta})}} \nonumber\\
 &\times 
 \left(
 u_{\beta}^\mu + \frac{W_{\beta}^\mu}{E_L + \frac{1}{3} E_{\beta} + \bm{\hat{W}}_{\beta} \cdot 
 \bm{\pi}_{\beta} \cdot \bm{\hat{W}}_{\beta}}\right),
 \label{eq:kg_uL}
\end{align}
where $W_{\beta}^2 = \rho^2\Omega^6 \gamma^{12} / 324 \beta_0^4 \ge 0$,
$\bm{\hat{W}}_{\beta} \equiv W_{\beta}/\sqrt{W_{\beta}^2}$
and $\bm{\hat{W}}_{\beta} \cdot \bm{\pi}_{\beta} \cdot \bm{\hat{W}}_{\beta} = 
\Omega^2 \gamma^6/ 54\beta_0^2$.
Surprinsingly, the Landau frame is well-defined only for 
$0 \le \rho\Omega \le (\rho\Omega)_{\rm lim}$, where
\begin{equation}
 (\rho\Omega)_{\rm lim} = \sqrt{x^2 + x + 1} - x,\qquad 
 x = \frac{5}{4\pi^2} (\beta_0\Omega)^2. 
\end{equation}
When $\rho\Omega > (\rho\Omega)_{\rm lim}$, $E_L$ is no longer real. It can be seen that 
$(\rho\Omega)_{\rm lim}$ decreases from $1$ at $\beta_0 \Omega = 0$ (large temperatures 
or slow rotation) down to $0.5$ as $\beta_0\Omega \rightarrow \infty$.

\begin{figure}
\begin{center}
\begin{tabular}{c}
\includegraphics[width=0.9\linewidth]{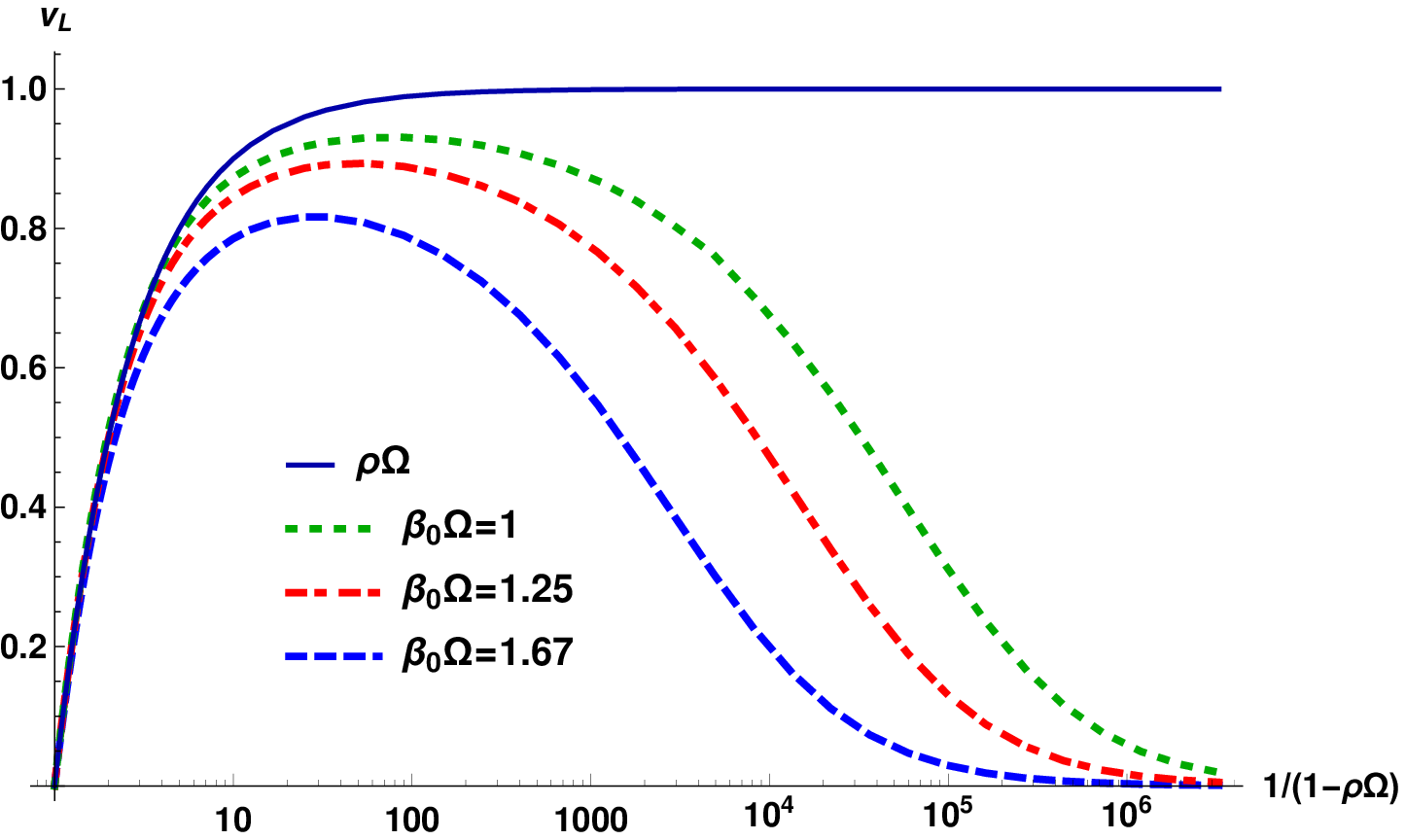} \\
(a)\\
\includegraphics[width=0.9\linewidth]{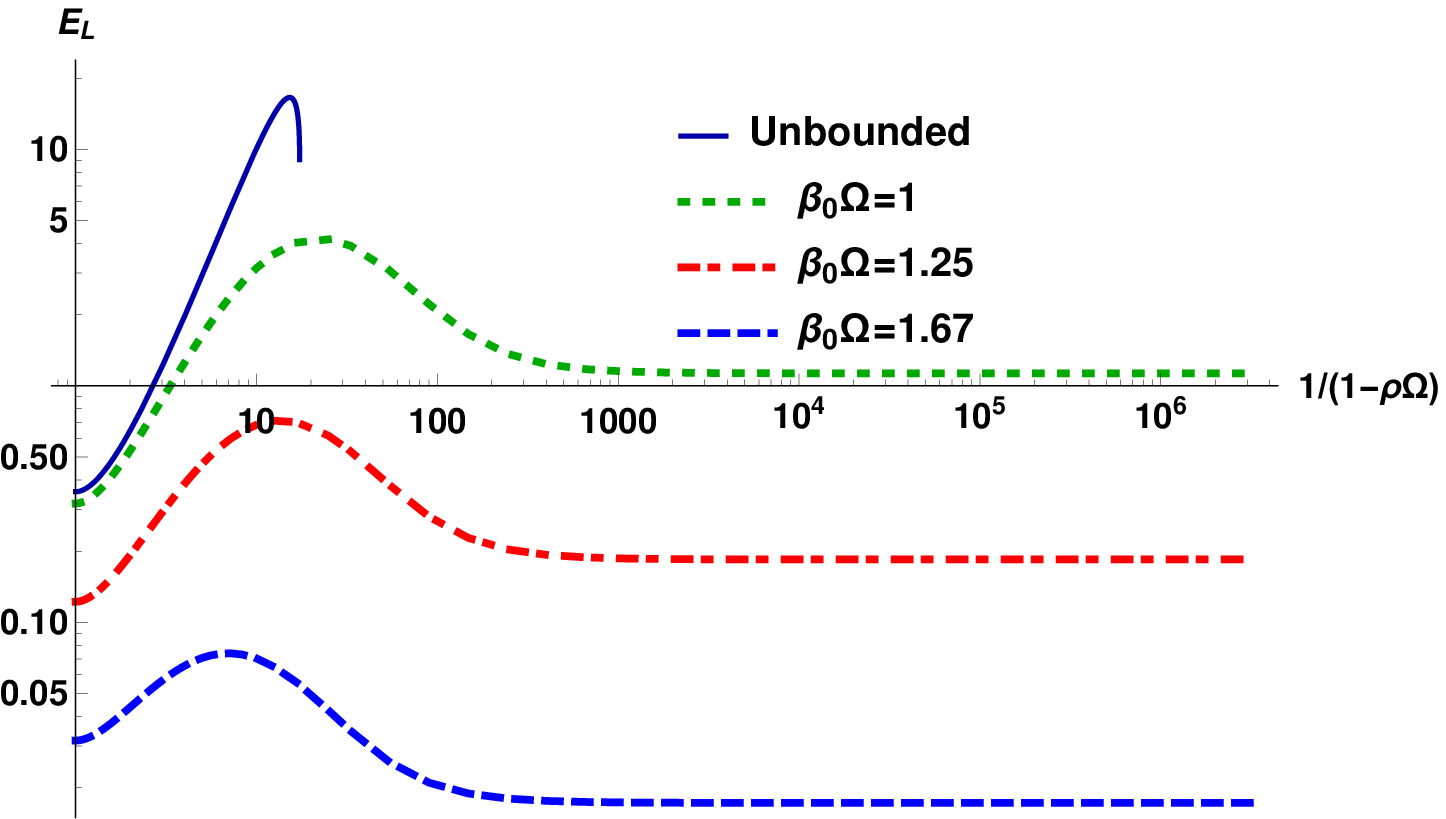} \\
(b) 
\end{tabular} 
\end{center}
\caption{(a) Landau velocity $v_L = \rho u_L^\varphi / u_L^0$ 
and (b) Landau energy $E_L$ for massless Klein-Gordon particles enclosed 
inside a cylinder located on the SOL ($R = \Omega^{-1}$). 
The continuous curve in (a) shows the velocity $\rho\Omega$ for the case of rigid rotation, while 
in (b) it corresponds to the Landau energy \eqref{eq:kg_EL} in the unbounded 
case. This curve is interrupted when $E_L$ becomes complex and the Landau frame 
is no longer well-defined.
}
\label{fig:kg}
\end{figure}

We are again forced to regard the RRTS of the Klein-Gordon field as ill-defined. 
The natural question to ask is whether the problem with defining the Landau frame persists when 
the system is enclosed inside a boundary. Following Ref.~\cite{duffy03}, 
we construct the Landau frame for the case when the system is enclosed inside 
a cylinder of radius $R = \Omega^{-1}$ (i.e.~placed on the SOL), on which Dirichlet boundary conditions 
are imposed. Fig.~\ref{fig:kg} shows that the Landau frame is well defined arbitrarily close to the boundary, 
where the Landau velocity $v_L = \rho u_L^\varphi / u_L^0$ decreases to $0$ due to the boundary conditions.
It can also be seen in Fig.~\ref{fig:kg} that both $v_L$ and $E_L$ increase monotonically as
$\beta_0$ is increased. Figure~\ref{fig:kg}(b) also shows $E_L$ for the unbounded Minkowski space 
\eqref{eq:kg_EL} for the case when $\beta_0\Omega = 1$. The curve is interrupted 
at $\rho\Omega \simeq 0.942$, where $E_L$ becomes complex.

\section{Dirac field} \label{sec:dirac}

\begin{figure}
\begin{center}
\begin{tabular}{c}
\includegraphics[width=0.9\linewidth]{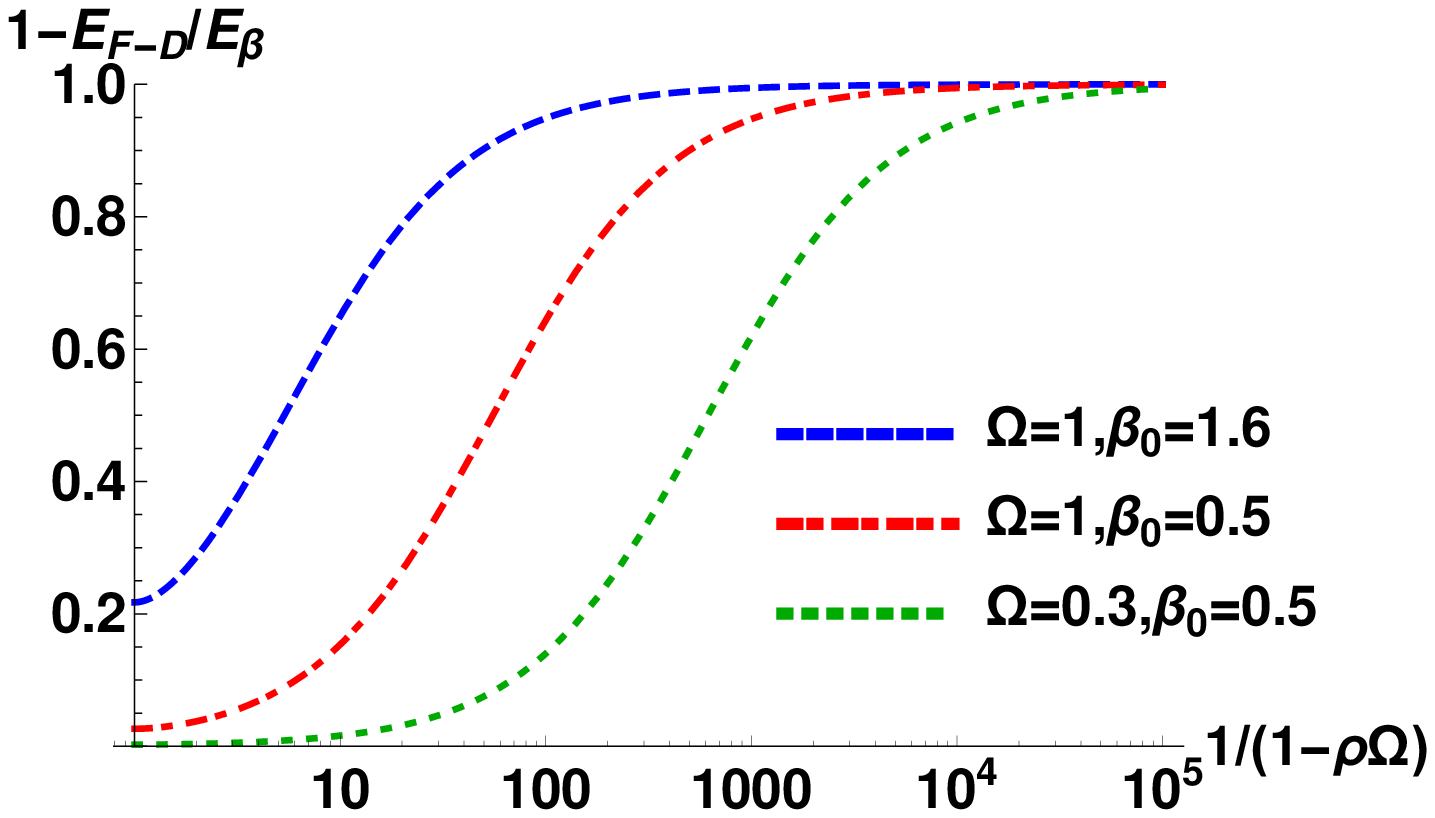} \\
(a) \\
\includegraphics[width=0.9\linewidth]{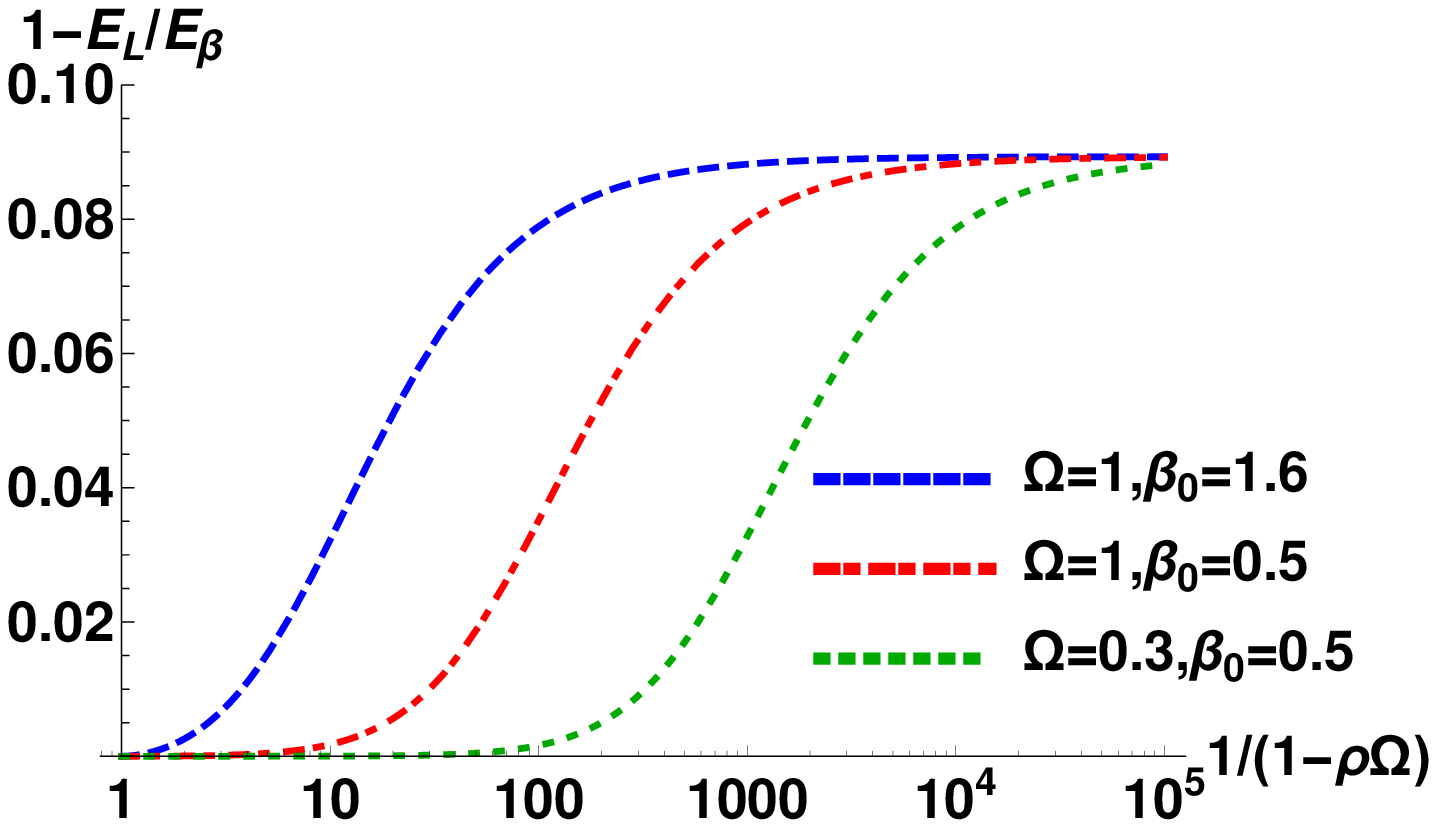}  \\
(b) \\
\includegraphics[width=0.9\linewidth]{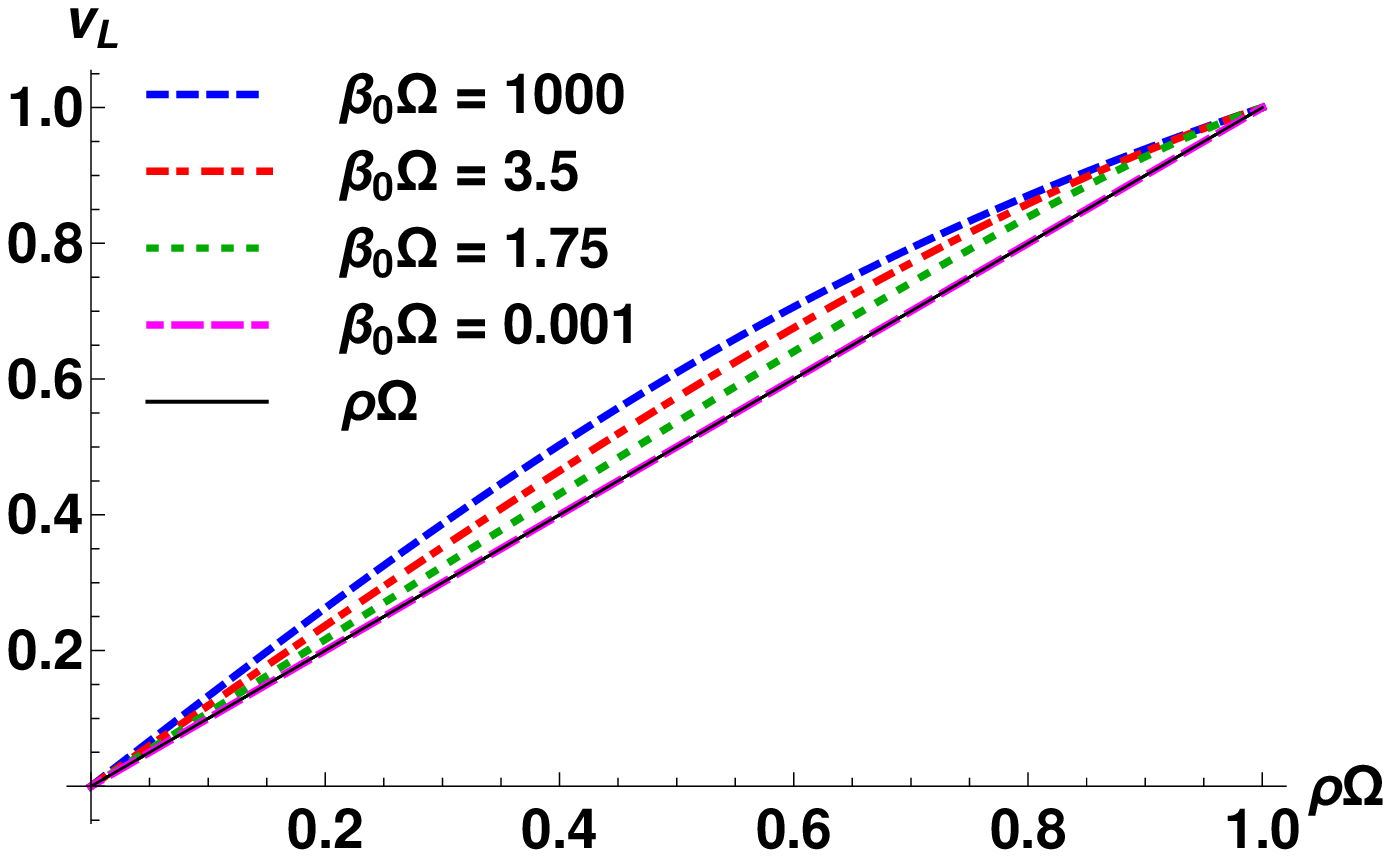} \\
(c) \\
\includegraphics[width=0.9\linewidth]{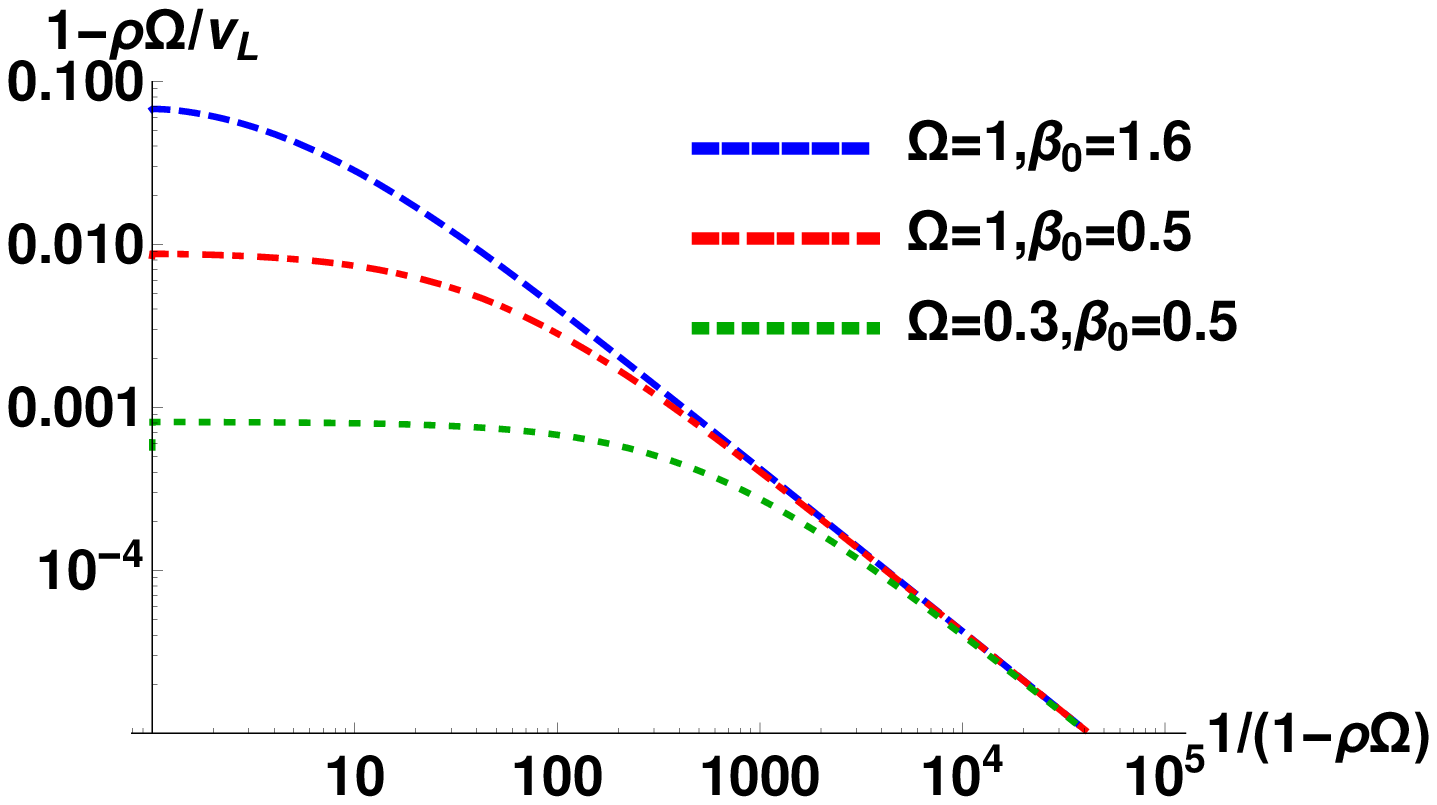} \\
(d)
\end{tabular} 
\end{center}
\caption{(a) Comparison between the energy density 
obtained from kinetic theory $E_{\rm F-D}$ \eqref{eq:E_kin} and the $\beta$-frame quantum energy 
density $E_{\beta}$ \eqref{eq:dirac_Ee}; (b) Comparison between the energy densities
$E_L$ \eqref{eq:dirac_EL} and $E_{\beta}$ \eqref{eq:dirac_Ee};
(c) Comparison between the Landau velocity $v_L = \rho u^\varphi_L / u_L^t$ and 
the velocity $\rho \Omega$ correponding to rigid rotation; 
(d) Relative difference $1 - \rho\Omega / v_L$ between $v_L$ and $\rho\Omega$ .
}
\label{fig:dirac}
\end{figure}

The QFT analysis of the RRTS of the Dirac field is presented in Ref.~\cite{ambrus14plb}.
The $\beta$-frame decomposition can be performed using 
$u_{\beta}$ \eqref{eq:beta_u} for the components of the SET given in Eqs.~(25c)--(25f) in 
Ref.~\cite{ambrus14plb}, yielding:
\begin{subequations}
\begin{align}
 E_{\beta} =& \frac{7\pi^2\gamma^4}{60 \beta_0^4} + \frac{\Omega^2}{24\beta_0^2} 
 \left(4\gamma^6 - \gamma^4\right),\label{eq:dirac_Ee}\\
 W_{\beta} =& \frac{\Omega^3 \gamma^7}{18\beta_0^2} (\rho^2\Omega \partial_t + \partial_\varphi),
 \label{eq:dirac_qe}
\end{align}
\end{subequations}
while $P_{\beta} = E_{\beta}/3$ and $\pi_{\beta}^{\mu\nu} = 0$. 
It is remarkable that $W^\mu_\beta$ for the Dirac field \eqref{eq:dirac_qe} has the same 
expression as that for the Klein-Gordon field \eqref{eq:KG_kin}.
As in the case of the Klein-Gordon field, 
the first term in Eq.~\eqref{eq:dirac_Ee} corresponds to $E_{\rm F-D}$ \eqref{eq:E_kin}, while the 
second term represents a quantum correction which dominates in the vicinity of the 
SOL. Figure~\ref{fig:dirac}(a) demonstrates this behaviour and it can be seen that the 
correction increases when either $\Omega$ or $\beta$ are increased.

The eigenvalue equation \eqref{eq:Landau_def} can be solved analytically in terms of the 
Landau energy and velocity:
\begin{align}
 E_L =& \frac{E_{\beta}}{3} + \sqrt{\frac{4E_{\beta}^2}{9} - W_{\beta}^2},\label{eq:dirac_EL}\\
 u_L^\mu =& \sqrt{\frac{3E_L + E_{\beta}}{2(3E_L - E_{\beta})}}
 \left(u_{\beta}^\mu + \frac{3W_{\beta}^\mu}{3E_L + E_{\beta}}\right).
\end{align} 
In contrast to the case of the Klein-Gordon field, the Landau frame is well-defined everywhere inside the SOL,
since $4E_{\beta}^2/9 W_{\beta}^2 > 1$ when $\rho\Omega < 1$. The ratio $E_L/E_{\beta}$ decreases from 
$1$ on the rotation axis down to $\frac{1}{3} + \frac{1}{\sqrt{3}}$ as the SOL is approached, where 
$W_{\beta} \rightarrow \frac{1}{3} E_{\beta}$. 
At fixed $\rho\Omega < 1$, $E_L$ approaches $E_{\beta}$ as either $\Omega$ or $\beta$ are decreased,
as confirmed in Fig.~\ref{fig:dirac}(b).

The Landau velocity $v_L = \rho u_L^\varphi / u_L^0 \ge \rho\Omega$ is compared to 
$\rho\Omega$ in Fig.~\ref{fig:dirac}(c). 
The difference $1 - \rho\Omega / v_L$ decreases to zero as the SOL is approached, while 
its value at the origin increases monotonically as $\beta_0 \Omega$ is increased.

For completeness, we list below $\pi^{\mu\nu}_L$:
\begin{multline*}
 \pi^{\mu\nu}_L =
 \frac{2(E_{\beta} -E_L)(3E_L - 2E_{\beta})}{3(3E_L - E_{\beta})} u_{\beta}^\mu u_{\beta}^\nu + 
 \frac{E_{\beta} - E_L}{3} g^{\mu\nu}\\
 - \frac{E_{\beta} - E_L}{3E_L - E_{\beta}} (u_{\beta}^\mu W_{\beta}^\nu + u_{\beta}^\nu W_{\beta}^\mu)
 - \frac{6E_L}{9E_L^2 - E_{\beta}^2} W_{\beta}^\mu W_{\beta}^\nu.
\end{multline*}

\section{Astrophysical application} \label{sec:astro}

Let us now apply our results in the context of the 
millisecond pulsar PSR J1748--2446ad reported in Ref.~\cite{hessels06}. Its pulse frequency 
is $\nu \simeq 716\,{\rm Hz}$, such that the SOL is 
located at a distance $\rho_{\rm SOL} = c / 2\pi \nu \simeq 66.685 {\rm km}$ 
from the rotation axis.
The typical surface temperature for a neutron star 
with characteristic age $\tau_c \ge 2.5 \times 10^7\,{\rm years}$ is 
$T_s \simeq 10^5\,{\rm K}$ \cite{glendenning97}. Its radius is 
$r_s \lesssim 16\,{\rm km}$ \cite{hessels06}, such that 
the temperature on the rotation axis can be extrapolated as $T_0 = T_s / \gamma_s \simeq 
9.7 \times 10^4\,{\rm K}$. Let us now investigate the magnitude of the quantum corrections for 
massless Dirac fermions dragged into rigid rotation by the pulsars magnetic field 
($B_{\rm surf} \le 1.1 \times 10^9\,{\rm G}$ \cite{hessels06}) by considering the folowing quantity:
\begin{multline}
 \delta E_{\rm QFT} = 
 \frac{E_{\beta}}{E_{\rm F-D}} - 1= \frac{10}{7\pi^2} 
 \left(\frac{h \nu}{K_B T_0}\right)^2 \left(\gamma^2 - \frac{1}{4}\right)\\
 \simeq 1.8 \times 10^{-26} \left(\gamma^2 - \frac{1}{4}\right),
\end{multline}
where the appropriate units were reinserted.
As pointed out in Ref.~\cite{becattini15}, the quantum correction is very small due to the 
presence of the Planck constant $h$. The value of $\gamma$ at which $E_{\beta} = 2 E_{\rm F-D}$
is $\gamma \simeq 7.4 \times 10^{12}$, which would correspond for an electron to an energy of
$m \gamma c^2 \simeq 3.8 \times 10^{18}\,{\rm eV}$, comparable to cosmic rays energies.
At such high values of $\gamma$, the distance to the SOL is of order 
$\sim 6 \times 10^{-22}\,{\rm m}$, where the rotation of the accretion disk is most likely no 
longer rigid.

Since our analysis was performed at the level of massless fermions, it is worth mentioning 
that in the case of the pulsar PSR J1748--2446ad, the relativistic coldness \cite{rezzolla13}
has the value $\zeta_0 = mc^2 / k_B T_0 \simeq 6.1 \times 10^4$ in the case of 
electrons, while the ratio $mc^2 / h\nu \simeq 1.7 \times 10^{17}$ also has a large value.
These numbers indicate that the massless limit results presented in this paper may be inaccurate
close to the rotation axis, where the properties of RRTS are heavily influenced by the value of $m$ 
in both the kinetic theory \cite{ambrus15wut} and in the QFT \cite{ambrus14plb} approaches. 
Also in these latter references, it can be seen that the mass dependence dissapears in the vicinity 
of the SOL, such that at $\gamma \sim 7.4 \times 10^{12}$, the particle constituents 
behave as though they were massless.

\section{Conclusion} \label{sec:conc}

In summary, we investigated rigidly-rotating thermal states of massless Klein-Gordon and 
Dirac particles. In comparison to relativistic kinetic theory results, the QFT approach yields a 
non-ideal SET. An analysis of the quantum SET reveals the presence 
of quantum corrections to the energy density, as well as non-equilibrium terms such as 
the shear pressure tensor.
These quantum terms become dominant as the speed of light surface (SOL) is approached. 
While for the Dirac field, the Landau frame can be defined everywhere up to the SOL, this is 
not so for the Klein-Gordon field, which we analysed based on the quantum corrections 
calculated in Ref.~\cite{becattini15}. The Landau frame becomes everywhere well defined 
when the system is enclosed inside a boundary placed inside or on the SOL.

An evaluation of the order of magnitude of the quantum corrections in a realistic 
astrophysical system (i.e. for a millisecond pulsar) shows that for such systems,
quantum corrections become important only at cosmic ray energies, in which case
the rigid rotation must be mantained up to subnuclear distances from the SOL.

\section*{Acknowledgements}
The author would like to thank Robert Blaga for preliminary discussions and for reading the manuscript.
This work was supported by a grant of the Romanian National Authority for Scientific Research and Innovation,
CNCS-UEFISCDI, project number PN-II-RU-TE-2014-4-2910. 

\appendix

\section{QFT analysis of the Klein-Gordon field}\label{app:kg}

It is well-known that the t.e.v.~of the SET for the RRTS of the Klein-Gordon (KG) field is ill-defined 
throughout the whole space-time \cite{ambrus14plb,duffy03}. It is also known that this anomalous behaviour 
is due to modes which are not present once the system is enclosed within a boundary which excludes the space 
outside of the SOL \cite{duffy03,nicolaevici01,vilenkin80}. Moreover, the kinetic theory treatment of the 
same system allows the SET to be computed uneventfully everywhere inside the SOL. 
Recently, quantum corrections to these kinetic theory results were reported in Ref.~\cite{becattini15}.
The purpose of this appendix is to bridge the gap between the perturbative analysis of Ref.~\cite{becattini15} 
and the expressions obtained from the exact QFT approach.

The QFT analysis of the RRTS of the KG field can be performed following Refs.~\cite{ambrus14plb,duffy03}
by introducing co-rotating coordinates $x^\mu_r = (t_r, \rho_r, \varphi_r, z_r)$, defined via 
$\varphi_r = \varphi - \Omega t$, such that:
\begin{equation}
 ds^2 = -\gamma_r^{-2} dt_r^2 + 2\rho^2_r \Omega \, dt_r \, d\varphi_r + d\rho_r^2 + 
 \rho_r^2 d\varphi_r^2 + dz_r^2.
 \label{eq:ds_rot}
\end{equation}
The KG field operator for scalar particles of mass $\mu$ can be expanded as:
\begin{equation}\label{eq:phi}
 \Phi(x_r) =  \sum_{m_j = -\infty}^\infty \int_{\mu}^\infty \omega_j \, d\omega_j \int_{-p_j}^{p_j} dk_j
 \left[f_j(x_r) a_j+ f^*_j(x_r) a^\dagger_j\right],
\end{equation}
where $f_j(x_r) \equiv f_{\omega k m}(x_r)$ are the mode solutions of the KG equation \cite{ambrus14plb,duffy03}:
\begin{equation}\label{eq:fkqm}
 f_{\omega k m}(x_r) = \frac{1}{\sqrt{8\pi^2 |\omega|}} e^{-i\omegat t_r + im\varphi_r + ikz_r}
 J_m(q\rho_r).
\end{equation}
In the above, $\omegat = \omega - \Omega m$ is the eigenvalue of the co-rotating 
Hamiltonian $H_r = i\partial_{t_r}$, while the transverse momentum $q$, 
longitudinal momentum $k$ and Minkowski energy $\omega$ satisfy $\omega = \sqrt{q^2 + k^2 + \mu^2}$,
with $p = \sqrt{\omega^2 - \mu^2}$ being the Minkowski momentum.
The one-particle operators 
$a_j$ and $a_j^\dagger$ satisfy the canonical commutation relations
$[a_j, a_{j'}^\dagger] = \delta(j,j')$, where 
\begin{equation}
 \delta(j,j') = \delta_{m_j,m_{j'}} \delta(k_j - k_{j'}) \frac{\delta(\omega_j - \omega_{j'})}{|\omega_j|}.
\end{equation}

Let us now consider the renormalised t.e.v. of the SET operator in the ``new improved'' \cite{candelas77} form 
corresponding to conformal coupling in Ref.~\cite{birrell82}:
\begin{multline}
 \braket{:T_{\mu\nu}:}_{\beta_0} =\\
 \braket{:\frac{1}{3} \{\phi_{;\mu}, \phi_{;\nu}\} - 
 \frac{1}{6} \{\phi, \phi_{;\mu\nu}\}
 -\frac{1}{6} g_{\mu\nu}(\phi^{;\lambda} \phi_{;\lambda} + \mu^2 \phi^2):}_{\beta_0},
\end{multline}
where the colon indicates normal (Wick) ordering. 
The anticommutator $\{,\}$ was introduced to esure operator symmetrisation.
The above t.e.v.~can be computed starting from \cite{ambrus14plb,vilenkin80}:
\begin{equation}
 \braket{:a_j a^\dagger_{j'}:}_{\beta_0} = \frac{\delta(j,j')}{e^{\beta_0\omegat_j} - 1}.
\end{equation}
Introducing the notation $G_{abc}$ through \cite{ambrus14phd}:
\begin{equation}
 G_{abc} = \frac{1}{\pi^2} \sum_{m = -\infty}^\infty \int_{\mu}^\infty \frac{d\omega}{e^{\beta_0 \omegat} - 1} 
 \int_0^p dk\, \omega^a q^b m^c J_m^2(q\rho),\label{eq:Gabc_def}
\end{equation}
the t.e.v. of $\phi^2$ and of the SET can be written as \cite{ambrus14phd}:
\begin{align}
 \braket{:\phi^2:}_{\beta_0} =& \frac{1}{2} G_{000},\label{eq:kg_phi2_Gabc}\\
 \braket{:T_{tt}:}_{\beta_0} =& \frac{1}{2} G_{200} + \frac{1}{24}\left(\frac{d^2}{d\rho^2} + 
 \frac{1}{\rho} \frac{d}{d\rho} \right) G_{000},\nonumber\\
 \braket{:T_{\rho\rho}:}_{\beta_0} =& \frac{1}{2} G_{020} - \frac{1}{2\rho^2} G_{002} + 
 \frac{1}{8} \left(\frac{d^2}{d\rho^2} + \frac{5}{3\rho} \frac{d}{d\rho} \right) G_{000},\nonumber\\
 \braket{:T_{\varphi\varphi}:}_{\beta_0} =& \frac{1}{2\rho^2} G_{002} - 
 \frac{1}{24} \left(\rho^2 \frac{d^2}{d\rho^2} + 3 \rho \frac{d}{d\rho} \right) G_{000},\nonumber\\
 \braket{:T_{zz}:}_{\beta_0} =& \frac{1}{2} (G_{200} - G_{020} - \mu^2 G_{000}) - 
 \frac{1}{24} \left(\frac{d^2}{d\rho^2} + \frac{1}{\rho} \frac{d}{d\rho} \right) G_{000},\nonumber\\
 \braket{:T_{t \varphi}:}_{\beta_0} =& -\frac{1}{2} G_{101}.\label{eq:kg_SET_Gabc}
\end{align}
The functions $G_{abc}$ \eqref{eq:Gabc_def} are clearly divergent due to the Bose-Einstein density of states 
factor $(e^{\beta_0\omegat} - 1)^{-1}$. In this section, 
we will present a procedure to 
isolate the regular part $G_{abc}^{\rm reg}$ of $G_{abc}$ 
by splitting $G_{abc}$ as follows:
\begin{equation}
 G_{abc} = G_{abc}^{\rm reg} + G_{abc}^{\infty},\label{eq:Gabc_split}
\end{equation}
where $G_{abc}^{\infty}$ absorbs the infinite part of $G_{abc}$.
We will show that $G_{abc}^{\rm reg}$ leads to the corrections presented in 
Ref.~\cite{becattini15}.

The method that we will employ is analogous to that used in Ref.~\cite{ambrus14plb} for Dirac fermions,
being based on expanding the Bose-Einstein density of states factor as follows \cite{ambrus14phd}:
\begin{equation}
 \frac{1}{e^{\beta_0(\omega - \Omega m)} - 1} = \sum_{n = 0}^\infty \frac{(-\Omega)^n}{n!} m^n \frac{d^n}{d\omega^n} 
 \left(\frac{1}{e^{\beta_0\omega} - 1}\right),\label{eq:expansion}
\end{equation}
Since the left hand side of the above expression has a pole at $\omega = \Omega m$,
the above expansion is not well defined when $\omega < \Omega m$. It is worth mentioning 
that the modes for which $\omegat < 0$ are no longer allowed when the system is enclosed
inside a boundary placed inside or on the SOL \cite{duffy03,nicolaevici01}.
Despite the fact that the modes with $\omegat< 0$ cannot be excluded from the mode sum 
in Eq.~\eqref{eq:Gabc_def}, we will show that the above procedure can still be used to 
recover the results in Ref.~\cite{becattini15}.

Substituting the expansion \eqref{eq:expansion} into Eq.~\eqref{eq:Gabc_def} yields:
\begin{multline}
 G_{abc} = \frac{1}{\pi^2} \sum_{n = 0}^\infty \frac{(-\Omega)^n}{n!}  
 \int_{\mu}^\infty d\omega \, \omega^a \frac{d^n}{d\omega^n} (e^{\beta_0\omega} - 1)^{-1} \\
 \times \int_0^p dk\, q^b \sum_{m = -\infty}^\infty m^{n+c} J_m^2(q\rho).\label{eq:Gabc_exp}
\end{multline} 
The sum over $m$ can be performed using the following formula:
\begin{equation}
 \sum_{m = -\infty}^\infty m^{2n} J_{m}^2(z) = \sum_{j = 0}^n \frac{\Gamma(j + \frac{1}{2})}{j! \sqrt{\pi}} a_{n,j} z^{2j},
\end{equation}
where the coefficients $a_{n,j}$ can be determined as follows:
\begin{equation}
 a_{n,j} = \frac{1}{(2j)!} \lim_{\alpha \rightarrow 0} \frac{d^{2n}}{d\alpha^{2n}} \left(2 \sinh \frac{\alpha}{2} \right)^{2j},
\end{equation}
such that $a_{n,j}$ vanishes when $j > n$. The following particular cases 
are required to evaluate Eqs.~\eqref{eq:kg_phi2_Gabc} and \eqref{eq:kg_SET_Gabc}:
\begin{gather}
 a_{j,j} = 1, \qquad a_{j+1, j} = \frac{1}{12} j(2j+1)(2j+2), \nonumber\\
 a_{j+2,j} = \frac{1}{1440} j(2j+1)(2j+2)(2j+3)(2j+4)(5j-1).
\end{gather}
Furthermore, the integral over $k$ in Eq.~\eqref{eq:Gabc_exp} can be performed using 
Eq.~(A.11) in Ref.~\cite{ambrus14plb}:
\begin{equation}
 \int_0^p dk\,q^\nu = \frac{\Gamma(\frac{\nu}{2} + 1) \sqrt{\pi}}{2\Gamma(\frac{\nu + 1}{2} + 1)} p^{\nu + 1}.
\end{equation}

Let us apply the above procedure for $G_{000}$, which reduces to:
\begin{multline}
 G_{000} = \frac{1}{\pi^2} \sum_{j = 0}^\infty 
 \frac{(\rho\Omega)^{2j}}{2j+1} \sum_{n = 0}^\infty \frac{\Omega^{2n} a_{n+j,j}}{(2n+2j)!} \\\times
 \int_\mu^\infty d\omega \,p^{2j+1} \frac{d^{2n+2j}}{d\omega^{2n+2j}} (e^{\beta_0\omega} - 1)^{-1}.
\end{multline}
In the massless case, $p =\omega$ and the integral over $\omega$ runs from $0$ to $\infty$. Noting that:
\begin{multline}
 \int_0^\infty d\omega\, \omega^{2j+1} \frac{d^{2n+2j}}{d\omega^{2n+2j}} (e^{\beta_0\omega} - 1)^{-1} = \\
 (2j+1)! \times
 \begin{cases}
  {\displaystyle \frac{\pi^2}{6\beta_0^2}},& n = 0,\\
  {\displaystyle -\frac{1}{2} + \frac{1}{\beta_0} \lim_{\omega \rightarrow 0} \omega^{-1}},& n = 1,\\
  {\displaystyle \frac{1}{\beta_0} (2n-2)! \lim_{\omega \rightarrow 0} \omega^{-2n+1}},& n > 1.
 \end{cases}
\end{multline}
It can be seen that the case $n =0$ corresponds to $G_{000}^{\rm reg}$.
The first term $-\frac{1}{2}$ in the $n = 1$ piece 
represents a temperature-independent contribution (i.e. which survives in the limit of vanishing 
temperature, when $\beta_0 \rightarrow \infty$).
This is the analogue of the spurious contributions highlighted in Ref.~\cite{ambrus14plb}, which 
are induced due to the construction of the thermal state with respect to the Minkowski (static) vacuum
(see the Iyer vs.~Vilenkin discussion in Ref.~\cite{ambrus14plb}). The second term in the $n = 1$ 
piece and all further terms with $n > 1$ are divergent, being induced by the infrared divergence of 
the Bose-Einstein density of states factor:
\begin{equation}
 \frac{1}{e^{\beta_0\omega} - 1} = \frac{1}{\beta_0 \omega} - \frac{1}{2} + \text{odd, positive powers 
 of } \beta_0 \omega.
\end{equation}
The result can be summarised as follows:
\begin{align}
 G_{000}^{\rm reg} =& \frac{\gamma^2}{6\beta_0^2}, \nonumber\\
 G_{000}^{\infty} =& -\frac{\Omega^2 \gamma^2}{24\pi^2}(\gamma^2 -1 )+ \frac{\Omega^2}{\pi^2 \beta_0} 
 \sum_{j = 0}^\infty (\rho\Omega)^{2j} \nonumber\\
 &\times \sum_{n = 0}^\infty \frac{\Omega^{2n} (2j)! (2n)! a_{n+j+1,j}}{(2n+2j+2)!} 
 \lim_{\omega\rightarrow 0} \omega^{-2n-1}.
\end{align}
After a similar analysis of the rest of the terms appearing in Eq.~\eqref{eq:kg_SET_Gabc}, the following 
regular contributions $\phi^2_{\rm reg}$ and $T_{\mu\nu}^{\rm reg}$ to $\braket{:\phi^2:}_{\beta_0}$ and 
$\braket{:T_{\mu\nu}:}_{\beta_0}$ can be obtained:
\begin{align}
 \phi^2_{\rm reg} =& \frac{\gamma^2}{6\beta_0^2},\\
 T_{tt}^{\rm reg} =& \frac{\pi^2 \gamma^4}{90\beta_0^4}(4 \gamma^2 - 1) + \frac{\Omega^2 \gamma^6}{36\beta_0^2} 
 (6\gamma^2 - 5),\nonumber\\
 T_{\rho\rho}^{\rm reg} =& \frac{\pi^2 \gamma^4}{90 \beta_0^4} + \frac{\Omega^2 \gamma^6}{36\beta_0^2},\nonumber\\
 \frac{1}{\rho^2} T_{\varphi\varphi}^{\rm reg} =& \frac{\pi^2 \gamma^4}{90 \beta_0^4}(4\gamma^2 - 3) + 
 \frac{\Omega^2 \gamma^6}{36\beta_0^2} (6\gamma^2 - 5),\nonumber\\
 T_{zz}^{\rm reg} =& \frac{\pi^2 \gamma^4}{90 \beta_0^4} - \frac{\Omega^2 \gamma^6}{36\beta_0^2},\nonumber\\
 \frac{1}{\rho} T_{t\varphi}^{\rm reg} =& -\rho\Omega \left[\frac{2\pi^2\gamma^6}{45\beta_0^4} + 
 \frac{\Omega^2\gamma^6}{18\beta_0^2}(3\gamma^2 - 1)\right].
\end{align}
Performing the $\beta$-frame decomposition with respect to $u_{\beta}$ \eqref{eq:beta_u} 
on the above expressions yields Eqs.~\eqref{eq:KG_kin}.

\end{document}